\documentclass[preprint,12pt]{elsarticle}
\usepackage{amssymb}
\journal{NIM A}  %Nuclear Instruments and Methods in Physics Research Section A}

\begin{document}

\begin{frontmatter}



\title{Nuclear Reactor Monitoring with Gadolinium-Loaded Plastic Scintillator Modules }


\author{Sertac Ozturk}

\address{Department of Physics, Tokat Gaziosmanpasa University, 60250, Tokat, Turkey}

\begin{abstract}
In this study, simulation-based design and optimization studies of a gadolinium-loaded segmented plastic scintillator detector are presented for monitoring applications of nuclear reactors in Turkey using antineutrinos. For the first time in the literature, a multivariate analysis technique is introduced to suppress cosmic background for such a reactor antineutrino detector. 
\end{abstract}


\end{frontmatter}


\section{Introduction}
\label{Intro}
When the neutrino was proposed for the first time by Wolfgang Pauli in 1930 \cite{pauli} based on measurement of beta decay, he said "I have done a terrible thing. I have postulated a particle that cannot be detected". He was wrong. After about 25 years, the first neutrino detection, which was emitted by a nuclear reactor, was recorded by Clyde Cowan and Frederick Reines in 1956 \cite{cowan}. The neutrinos play an important role for better understanding of the Universe. That's why there are many ongoing and planned neutrino experiments worldwide. 

While the neutrinos come from different sources, nuclear reactors are human-controlled very intense antineutrino sources. Each fission process releases 6 $\overline{\nu_e}$, and a 1 GW nuclear reactor emits about $2 \times 10^{20}$ $ \overline{\nu_e}$ per second. This makes nuclear reactors great tools for neutrino physics research, and so there are many neutrino oscillation experiments located near nuclear reactors \cite{kamland, chooz, dayabay, reno, dchooz, juno}.

The neutrinos interact with matter with very low cross section, and  emitted antineutrino flux from the nuclear reactor core is used for reactor monitoring since the thermal power is directly related to antineutrino flux. Measured antineutrino event rate ($N_\nu$) can be expressed as $N_\nu = \gamma (1 + k) P_{th}$, where $P_{th}$ is the reactor's thermal power, $k$ is the time dependent factor that consider the time evolution of the fuel composition, and $\gamma$ is a constant, which depends on the detector properties \cite{korovkin}. Measurement of antineutrino flux and energy spectrum with a detector that is located outside the reactor buildings provide quasi real time information of reactor thermal power, operational status and fissile content. So, International Atomic Energy Agency (IAEA) has recommended development and utilization of antineutrino monitoring with compact detectors as part of nuclear safeguard activities \cite{IAEA}.

The first nuclear reactor construction in Turkey has recently started at Akkuyu and its first power unit with a thermal capacity of 3.2 \textit{GWt} is planned to start operation in 2023. Monitoring Akkuyu nuclear power plant using antineutrino flux measurement is an important nuclear safety issue. For this reason, design and optimization of a water Cherenkov detector was  done\cite{Ozturk}. In this paper, segmented detector structure with gadolinium(Gd)-loaded plastic scintillators to obtain higher background rejection is considered and optimization studies are performed using GEANT4 simulation toolkit\cite{Geant4}. In addition, for the very first time, a multivariate analysis technique is introduced for cosmic background suppression. 


\section{Detector Design}
\label{detector}
An antineutrino can be detected via the so-called inverse beta decay (IBD), resulting from charged-current antineutrino-proton scattering in the plastic scintillators: $ \overline{\nu_e} + p \rightarrow e^+ + n $. The energy threshold of this process is 1.8 \textit{MeV} and it has the highest reaction cross section considering the other possible reactions of  $ \overline{\nu_e} + d \rightarrow e^+ + n + n $ and $ \overline{\nu_e} + e^- \rightarrow  \overline{\nu_e} + e^- $. 

The outgoing positron emits two gamma rays by annihilation ($e^+ + e^- \rightarrow 2 \gamma $) and it produces a prompt signal. The subsequent second signal  is generated by the thermal neutron capture processes in the gadolinium-loaded plastic scintillator and gamma ray cascades, which have the total energy of  $\sim$2 \textit{MeV} and $\sim$8 \textit{MeV}, are produced with the following processes\cite{Oguri}:
\\
\\

$ n + p \rightarrow d +  \gamma$'s (2 \textit{MeV})

$ n + ^{155}Gd \rightarrow  ^{156}Gd +  \gamma$'s ( $\sim$ 8.6 \textit{MeV})

$ n + ^{157}Gd  \rightarrow  ^{158}Gd +  \gamma$'s (8 \textit{MeV})
\\
\\
The delayed time coincidence  between the prompt and subsequent signals (about 5-50 $\mu s$)  is used for triggering an antineutrino event. 

The nuclear reactor monitoring using antineutrino was recommended by the IAEA for nuclear safeguarding and so, there are many studies and experiments about reactor antineutrino measurements. Generally, Gd-doped liquid scintillator or Gd-doped water are proposed as an active medium for antineutrino detection. Using a solid state detector instead of a water-based one has some advantages. The first one is that its transportation is easier compared to the liquid-state and it gives more mobility. In addition, a solid-state detector prevents liquid-leakage problems and the danger of fire from the oil-based liquid scintillator. The second and more important advantage is that a solid state antineutrino detector can be divided into many segments easier than liquid-state, and energy deposition in each module could be measured separately. This approach provides a great signal - cosmic background separation. 

\begin{figure} []
 \centering
  \includegraphics[width=10 cm]{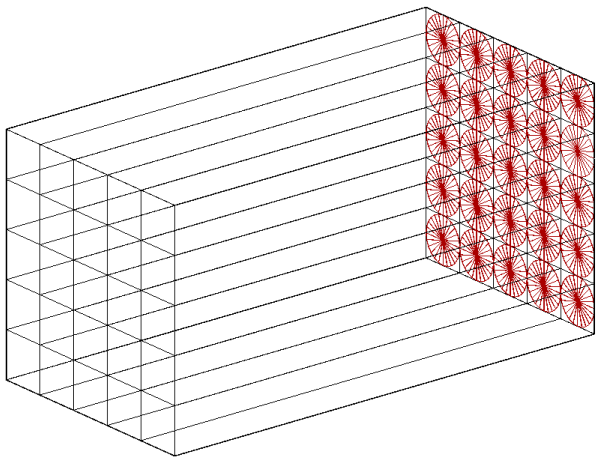}
  \caption{A schematic view of the detector. The circles present photo multiplier tubes (PMTs).}
  \label{fig:detector}
\end{figure}

For the remainder of this note, a segmented detector structure with gadolinium-loaded polyvinyltoluene based plastic scintillator is considered. A schematic view of the detector is shown in Figure~\ref{fig:detector}. The detector consists of 25 identical 10$\times$10$\times$100 \textit{cm} Gd-loaded plastic scintillators. Each plastic scintillator is wrapped in 20 $\mu m$ thick aluminium sheet to obtain a segmented structure. The reflectivity constant of aluminium sheet is set to 1. The weight of the designed detector is about 250 \textit{kg} and about 1185 antineutrino events can be observed per a day when it is placed 50 \textit{m} away from the 3.2 \textit{GWt} reactor core.

In previous studies\cite{Brudanin, Poehlmann}, Gd-loaded plastic scintillators were successfully synthesized and produced, and its effects on neutron capture process were investigated. These studies showed that transparency and the other optical properties of Gd-loaded scintillator with 1\%-3\% loading were almost the same as unloaded case. 

\begin{figure} []
 \centering
  \includegraphics[width=10 cm]{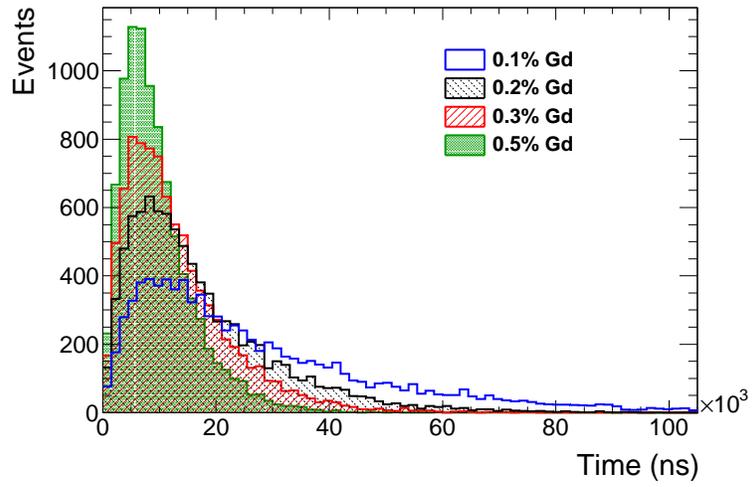}
  \caption{Time difference between prompt and delayed signals for various Gd concentrations.}
  \label{fig:gd}
\end{figure}

The amount of loaded Gd concentration in the scintillator directly effects the delayed signal, which is generated by thermal neutron capture. In this study, neutrons with 10 \textit{keV} energy were generated inside the detector using GEANT4 simulation toolkit and neutron capture time after production was investigated with the different Gd concentrations. Since positron decays immediately by annihilation after IBD, neutron capture time sets the delay between two signals. Figure~\ref{fig:gd} shows the simulation results of time differences between prompt and delayed signals for different amounts of Gd concentration. It was found that using plastic scintillator blocks with 0.2\%-0.3\% amount of Gd was optimum, which gave a prompt-delayed time difference between 4 and 50  $\mu s$, and amount of Gd concentration in the plastic scintillators was chosen 0.2\% for the remainder of theses studies. 

\begin{figure} [!h]
 \centering
  \includegraphics[width=10 cm]{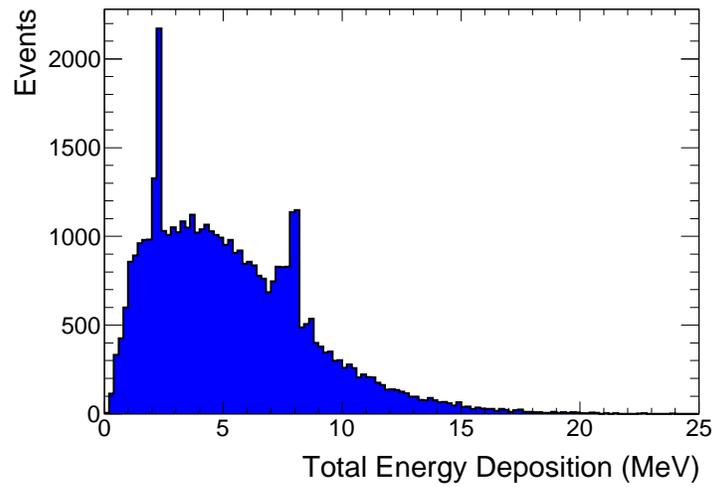}
  \caption{Simulation result of total energy deposition of neutrons with 10 \textit{keV} energy inside the detector.}
  \label{fig:energy}
\end{figure}

A gamma cascade is generated by thermal neutron capture process inside the detector medium. Simulation result of the total energy deposition in the detector is presented in Figure~\ref{fig:energy}. The thermal neutron capture by hydrogen gives a peak around 2 \textit{MeV}, and the neutron thermalization by $^{158}Gd$ and $^{156}Gd$ is observed to generate peaks around 8 \textit{MeV} and 8.6 \textit{MeV}, respectively. 

\clearpage

\section{Background and Multivariate Analysis}
\label{mva}
\subsection{Charged Cosmic Background}
The delayed coincidence of prompt-delayed signals is used as a trigger for antineutrino events. In addition, the segmented structure of the detector gives great separation between IBD candidate events and cosmic background. Since the antineutrino events and cosmic ray events have different event topology\cite{Kiff, Kiff2}, the energy correlation between PMT signals might be used for selecting the antineutrino events. 

\begin{figure} [!h]
 \centering
  \includegraphics[width=4 cm]{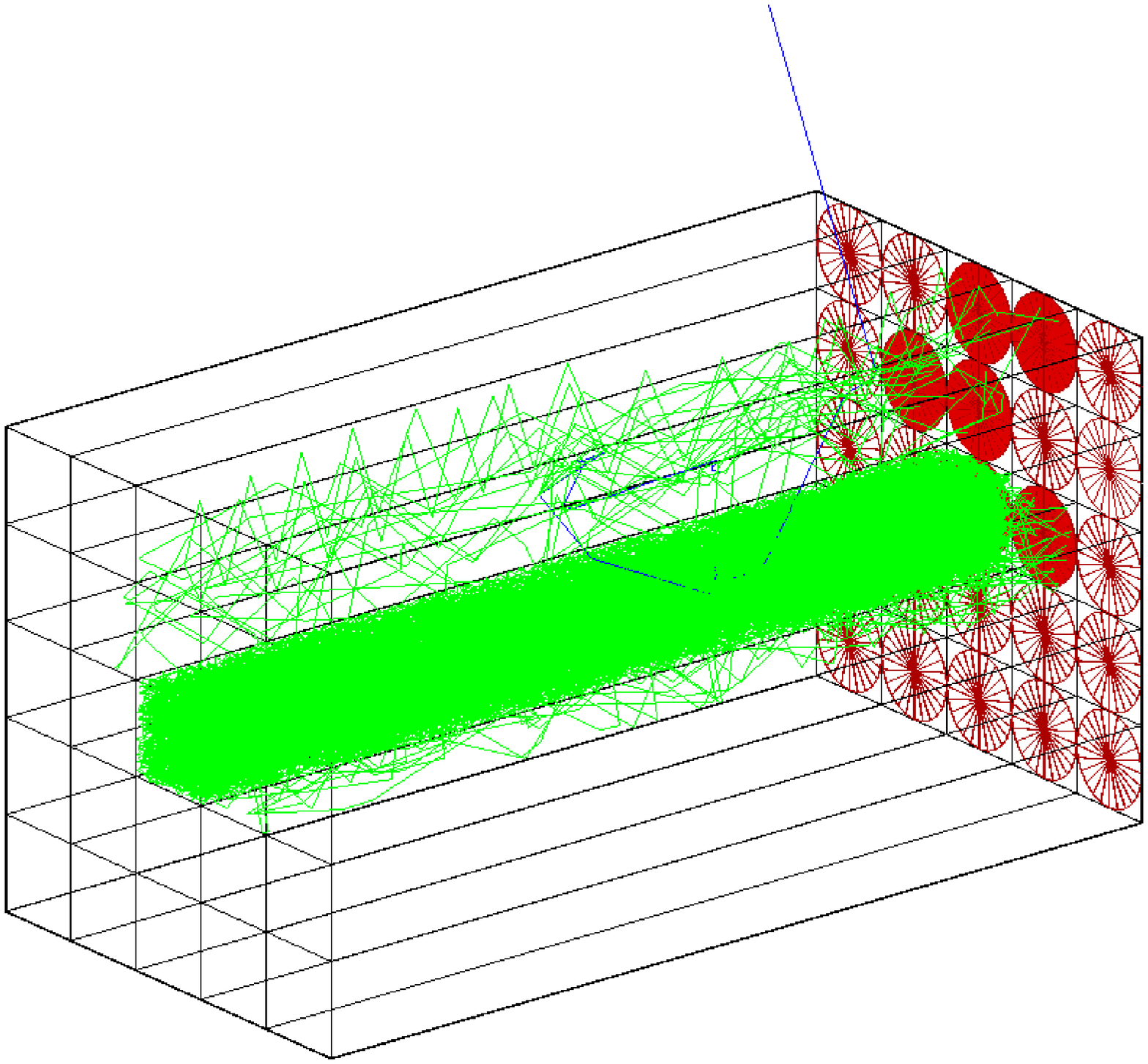}
  \includegraphics[width=4 cm]{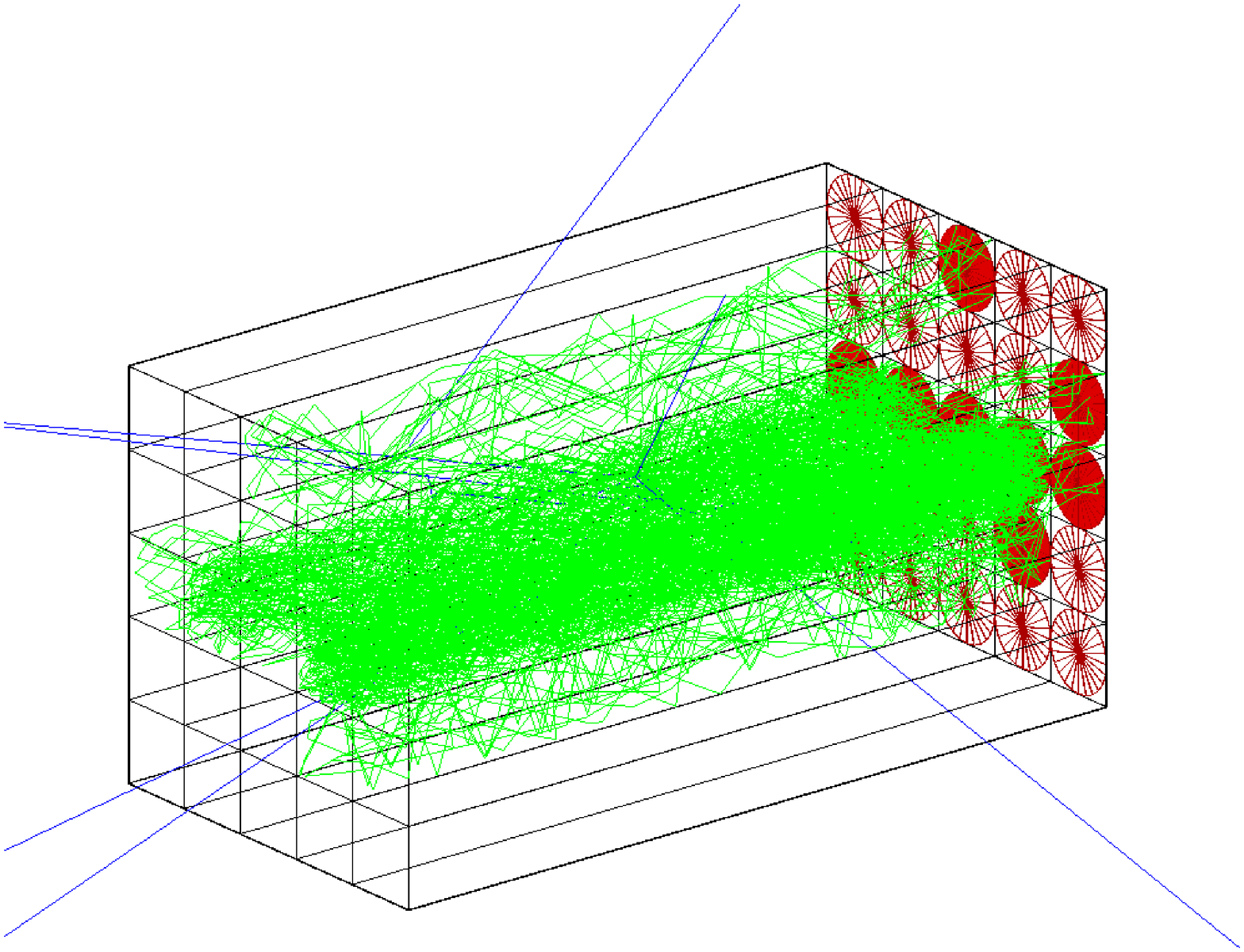}
  \includegraphics[width=4 cm]{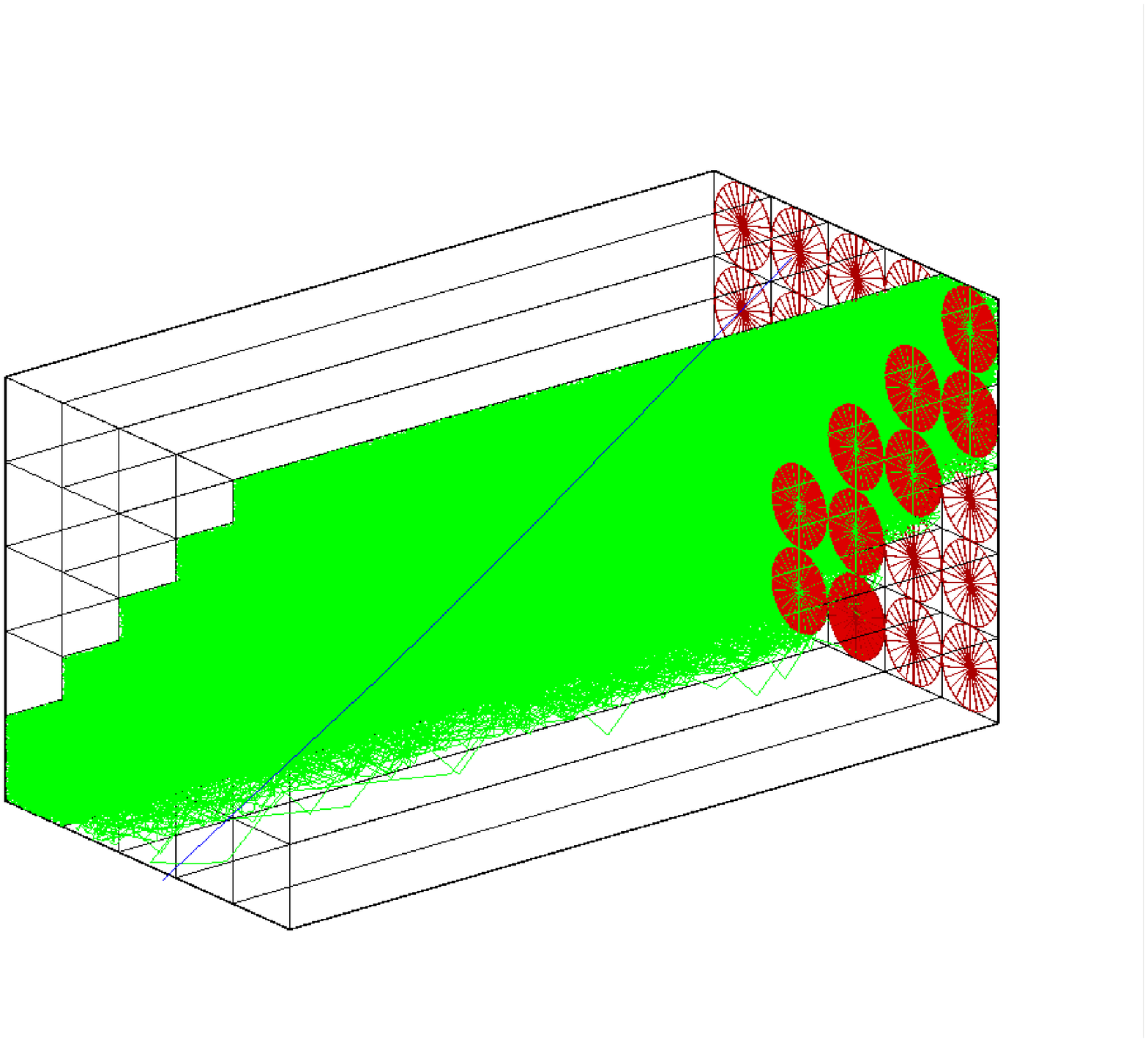}
  \caption{Events simulation of 3 \textit{MeV} positron (left) and 10 \textit{keV} neutron (middle) inside the  detector resulting of IBD are shown. Event simulation of a cosmic muon with the momentum of 1 \textit{GeV/c} passing though to the whole detector (right) is also presented.}
  \label{fig:topology}
\end{figure}

Figure~\ref{fig:topology} shows the GEANT4 simulation of an antineutrino like event and a cosmic muon event. Since antineutrino-proton scattering process was not available in GEANT4 (version 10.5)  libraries, positrons with the energy of 3 \textit{MeV} and neutrons with the energy of 10 \textit{keV} were generated separately in the center of detector medium. Muon, electron, positron and proton beams with 1 \textit{GeV/c} momentum passing through the whole detector were considered as charged cosmic background events.  As it is seen in Figure~\ref{fig:topology}, the number of photoelectron (PE) correlations between PMTs are expected quite to be different.

%Positron decays by annihilation immediately after its production and almost all event energy is read by a single PMT. 

Since the number of PE emitted by PMT's photocathode is proportional to the energy given by the PMT, energy term is used instead of PE counts. $E_{1st}$, $E_{2nd}$ and $E_{3rd}$ are the highest, the second-highest and the third-highest energy deposits among all the modules, respectively. $ E_{Total}$ is the total energy deposit in the whole detector. 

 \begin{figure} [!h]
  \includegraphics[width=7.5 cm]{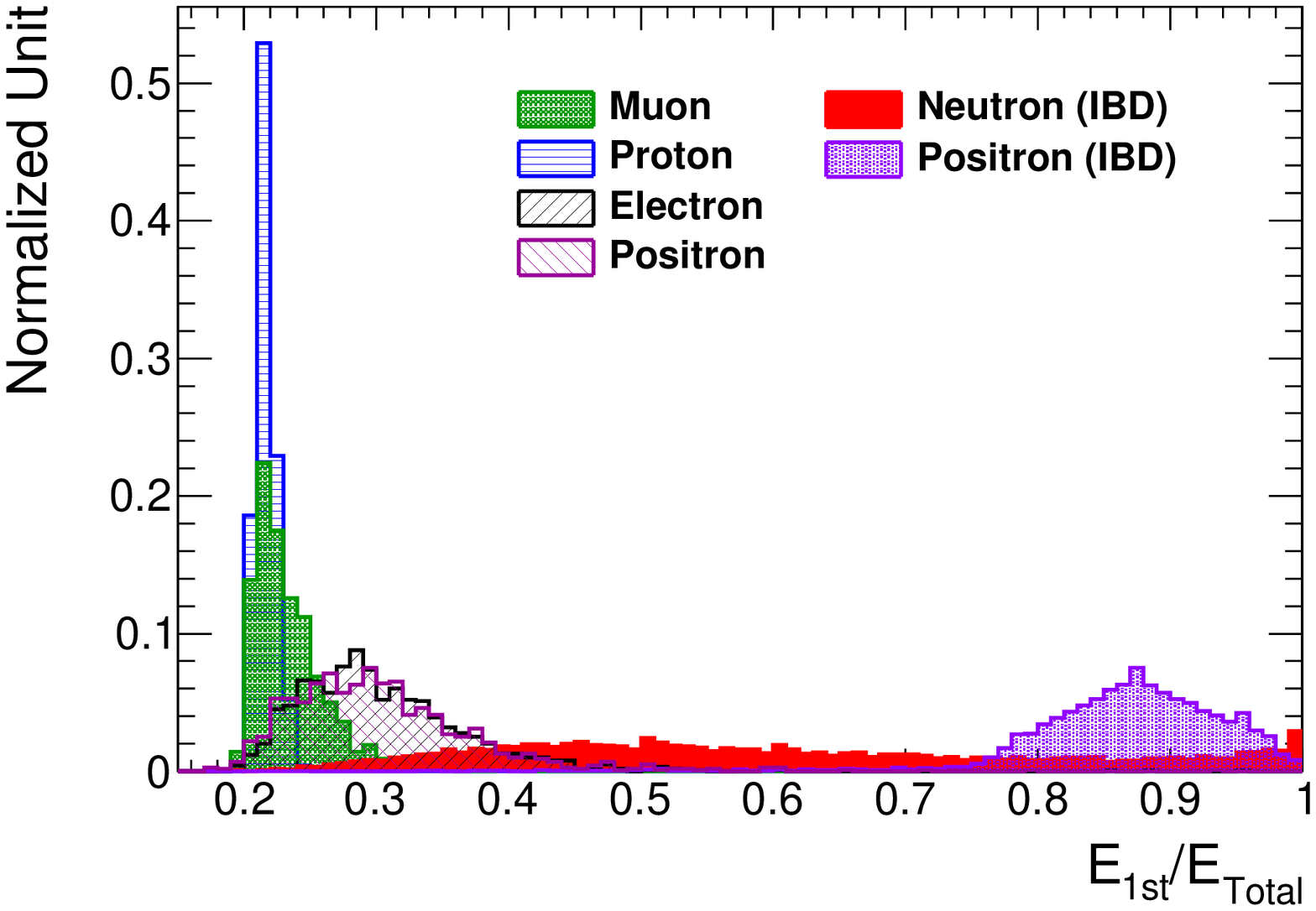}
  \includegraphics[width=7.5 cm]{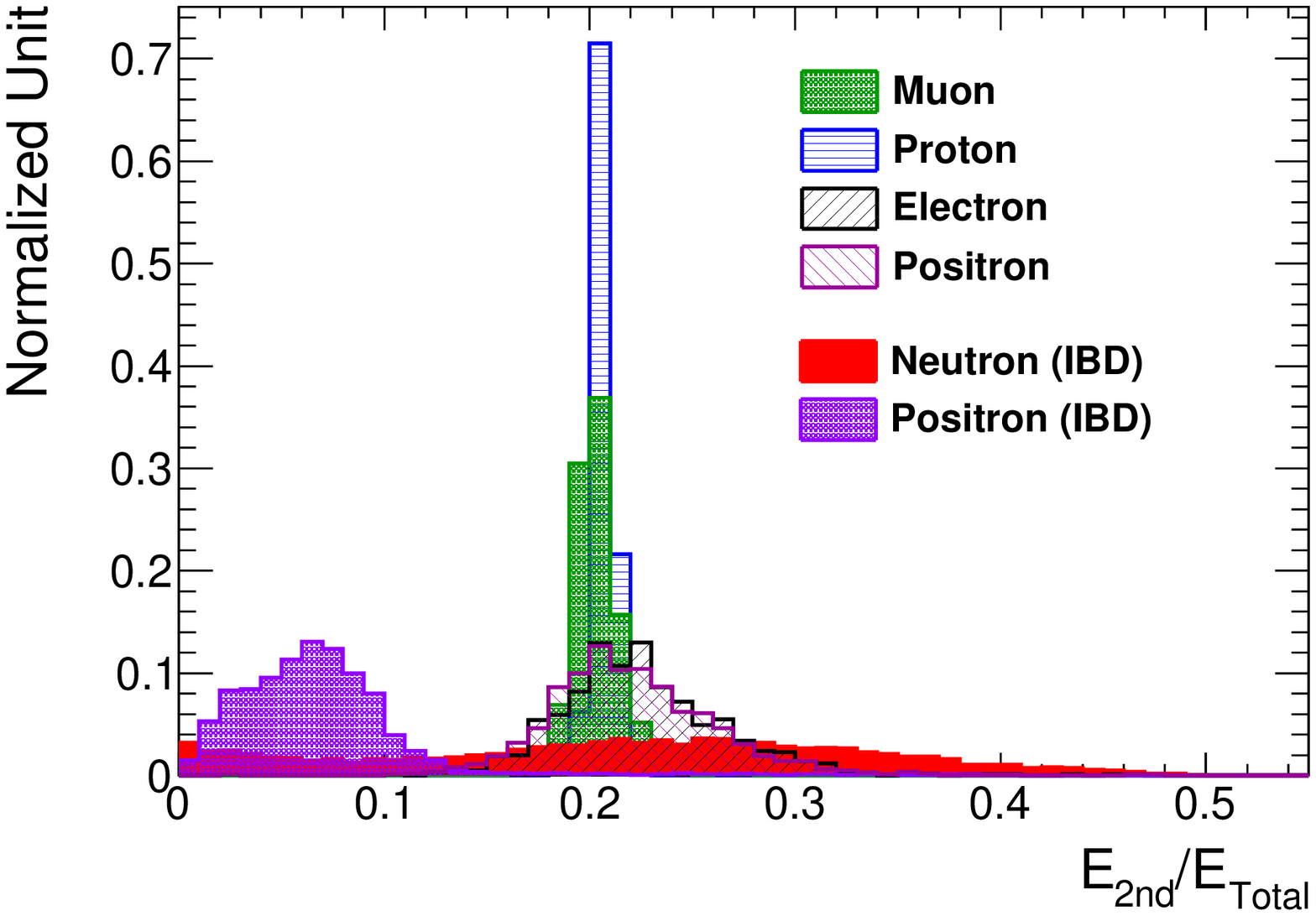}
  \includegraphics[width=7.5 cm]{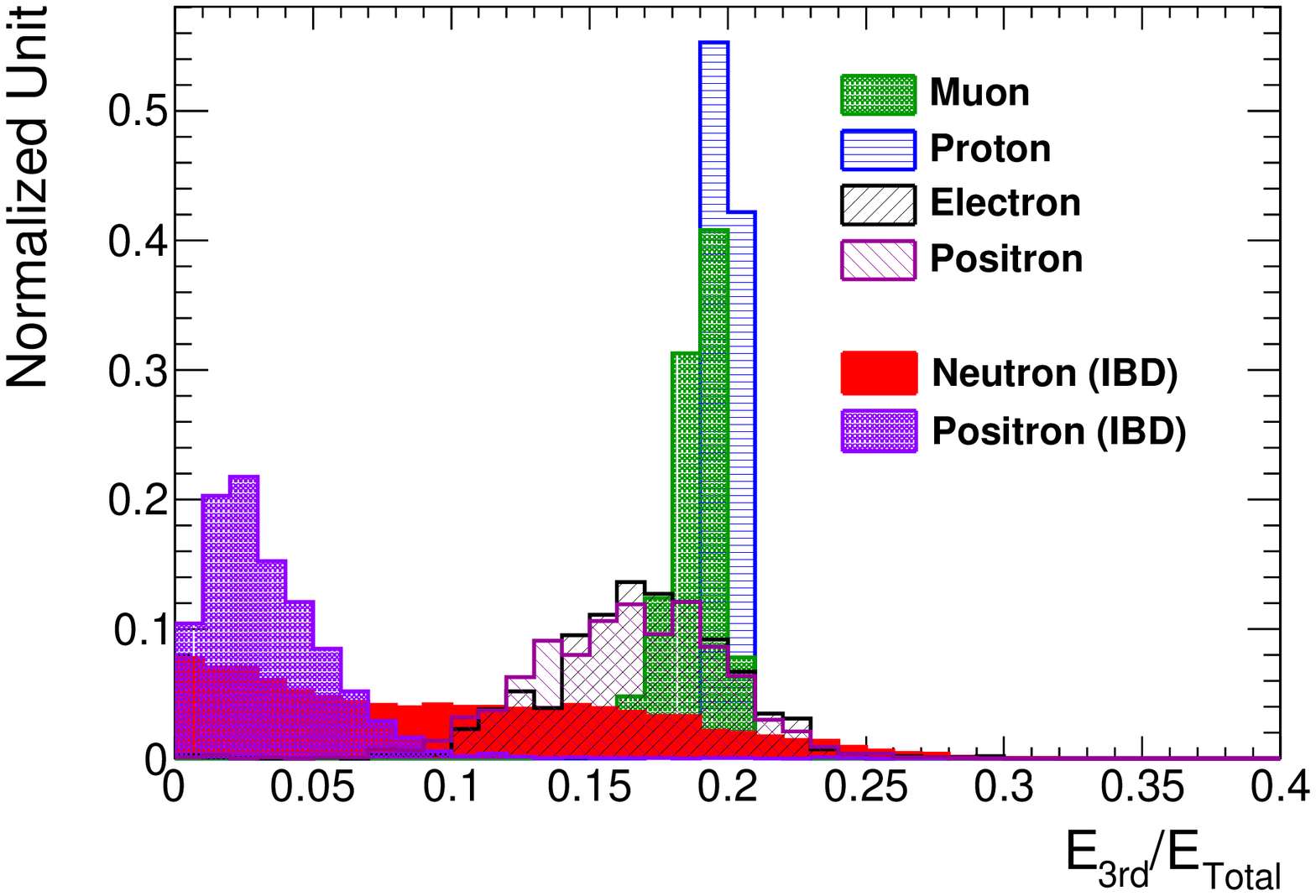}
  \caption{$E_{1st}/E_{Total}$, $E_{2nd}/E_{Total}$ and $E_{3rd}/E_{Total}$ distributions of antineutrino event like and charged cosmic particles events.}
  \label{fig:Eratio}
\end{figure}

$E_{1st}/E_{Total}$, $E_{2nd}/E_{Total}$ and $E_{3rd}/E_{Total}$ distributions of antineutrino events and charged cosmic rays events are shown in Figure~\ref{fig:Eratio}. $E_{1st}/E_{Total}$ is around 0.9 for 3 MeV positron that is initiated by IBD because it decays by annihilation immediately after its production and huge amount of the energy of the event is deposited in a single module. Charged cosmic particles with the momentum of 1 \textit{GeV/c} passing through the whole detector deposit energy homogeneously in the modules, and the first three highest energy fractions are around 0.2. $E_{1st}/E_{Total}$ distribution for thermal neutron capture process is roughly flat.

\begin{figure} [!h]
 \centering
  \includegraphics[width=6.5 cm]{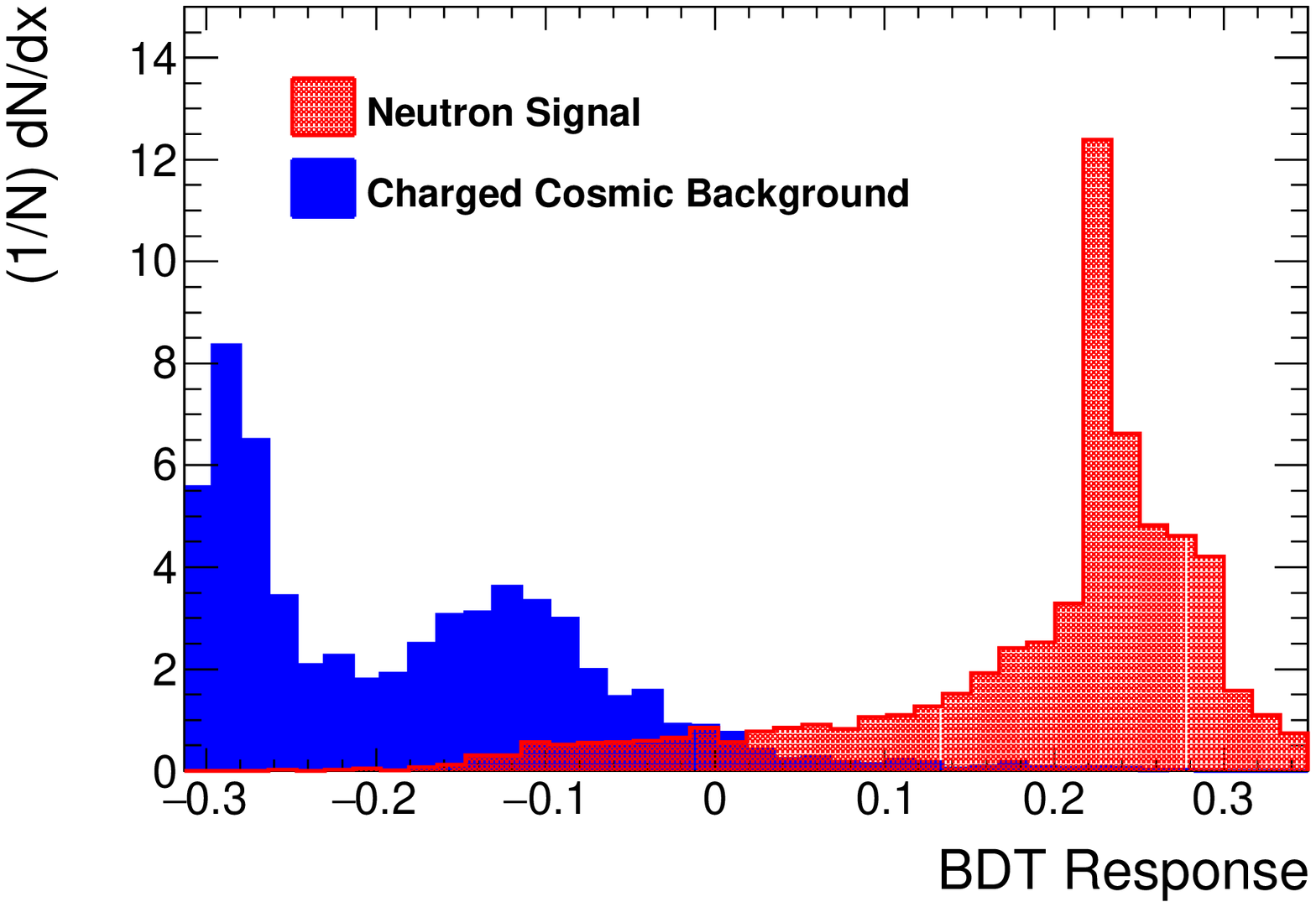}
   \includegraphics[width=6.5 cm]{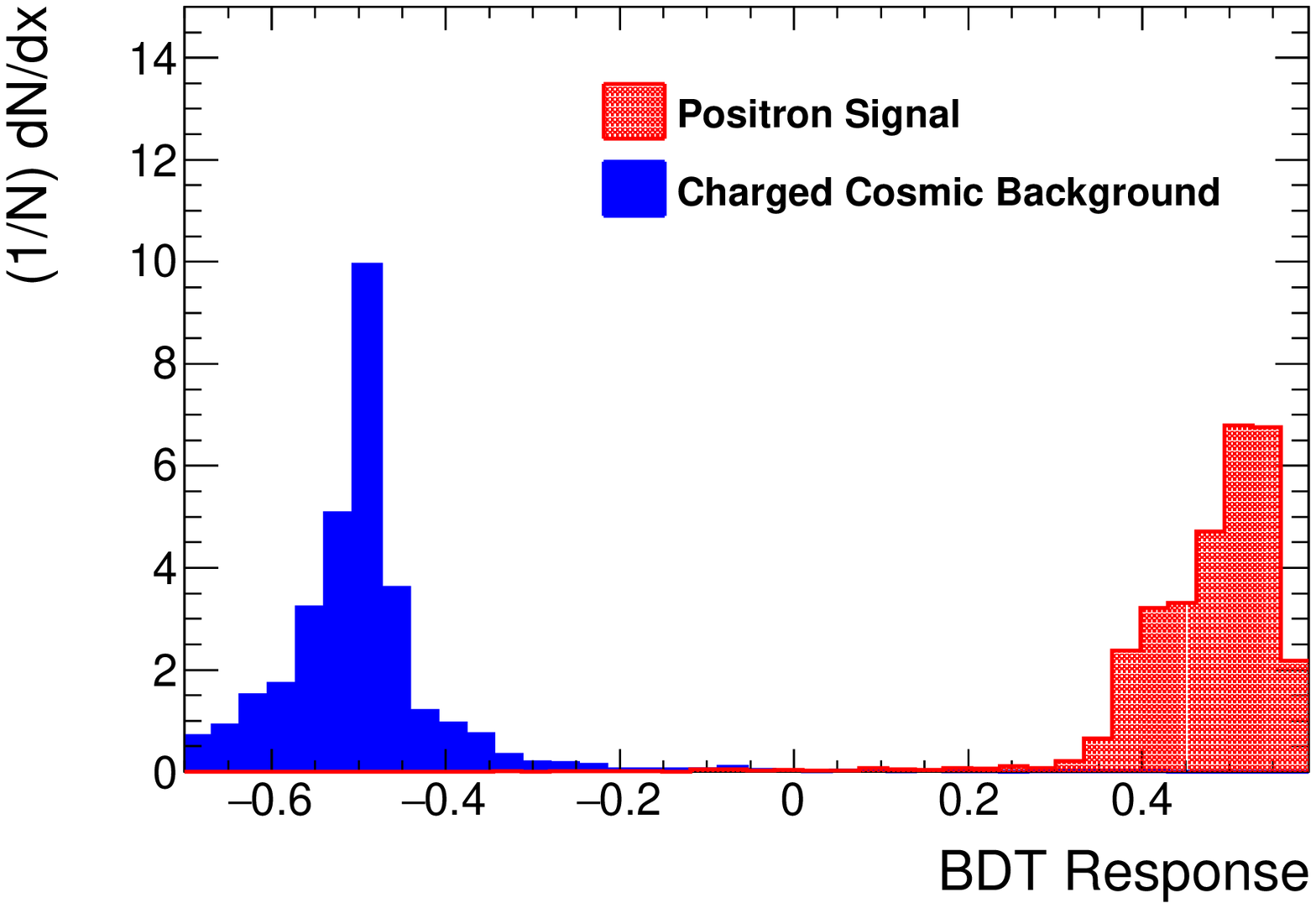}
  \caption{Boosted decision tree output distributions. Signal events are produced by neutron (left) and positron (right) resulting from IBD and the charged cosmic particles passing through the whole detector are considered as background events.}
  \label{fig:BDT_cosmic}
\end{figure}

As seen in Figure~\ref{fig:Eratio}, the highest energy fraction amongst all the modules ($E_{1st}/E_{Total}$) could be used to reject  background events. Generally, using several variables at the same time could improve background rejection significantly. In order to define a multivariate discriminant, the Toolkit for Multivariate Data Analysis  TMVA\cite{TMVA} part of ROOT v6.14.06\cite{Root} is used to combine $E_{1st}/E_{Total}$, $E_{2nd}/E_{Total}$ and $E_{3rd}/E_{Total}$ distributions. Boosted Decision Trees (BDT) method is chosen as a multivariate discriminator because it has given greater background rejection compared to Likelihood or Artificial Neural Networks methods. BDT is a machine-learning technique, widely used in experimental high energy physics analysis, that is the sequential application of cuts splits the data into nodes, where the final nodes (leaves) classify an event as signal or background\cite{TMVA}. The output distributions of BDT response for neutron and positron signals are shown in Figure~\ref{fig:BDT_cosmic}, respectively. Background events are taken as the sum of all considered cosmic particles as shown is Figure~\ref{fig:Eratio}, while the signal events are produced by IBD. There is a clear separation between signal and background events in BDT response distribution and a discriminant value on BDT response could be used event-by-event for charged cosmic background rejection.

\begin{figure} [!h]
\centering
  \includegraphics[width=10 cm]{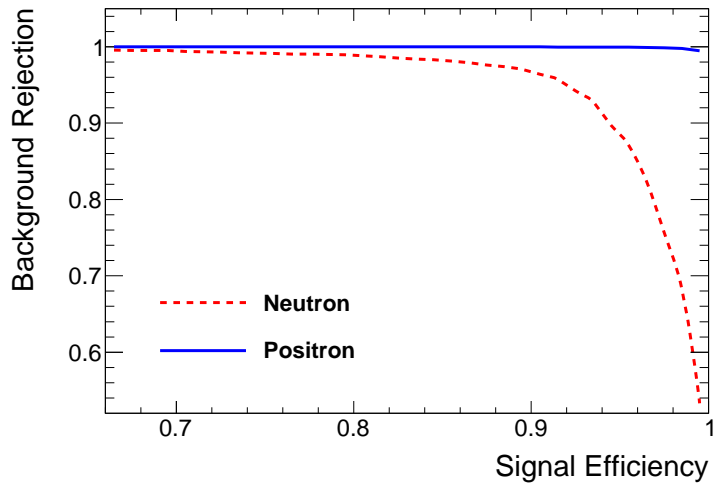}
  \caption{ROC curves for neutron and positron signals. The charged cosmic particles passing through the whole detector are considered as background events.}
  \label{fig:roc}
\end{figure}

The receiver operating characteristic (ROC) curves that describe the signal efficiency versus background rejection are shown in Figure~\ref{fig:roc}. It is seen that about 90\% of events, which are produced by high energetic charged cosmic particles passing through whole detector, could be rejected only considering energy deposition fractions in the modules while keeping around 95\% of antineutrino events.

\subsection{Neutron and Gamma Rays Background}
\begin{figure} [!h]
  \includegraphics[width=7.5 cm]{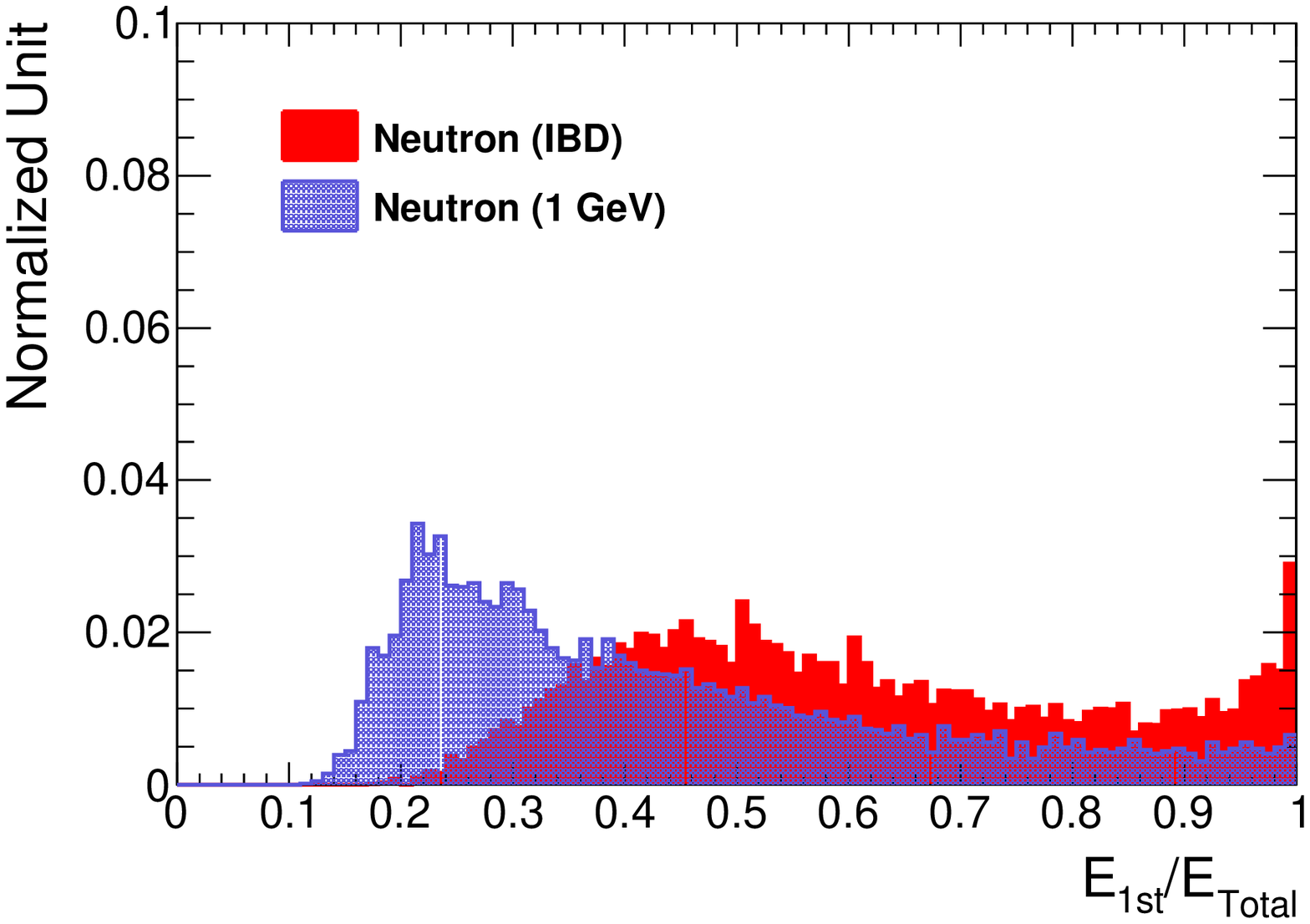}
  \includegraphics[width=7.5 cm]{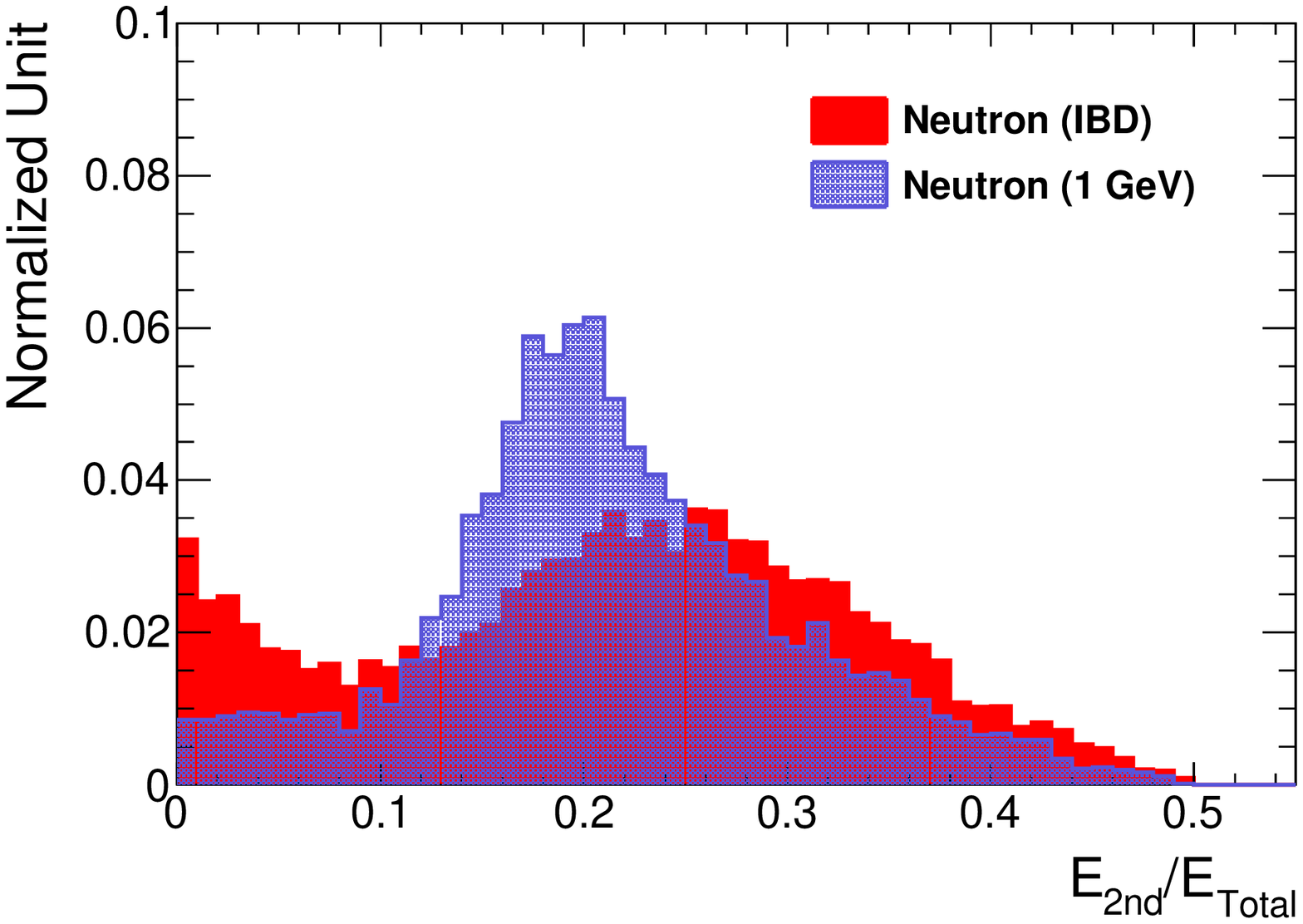}
  \includegraphics[width=7.5 cm]{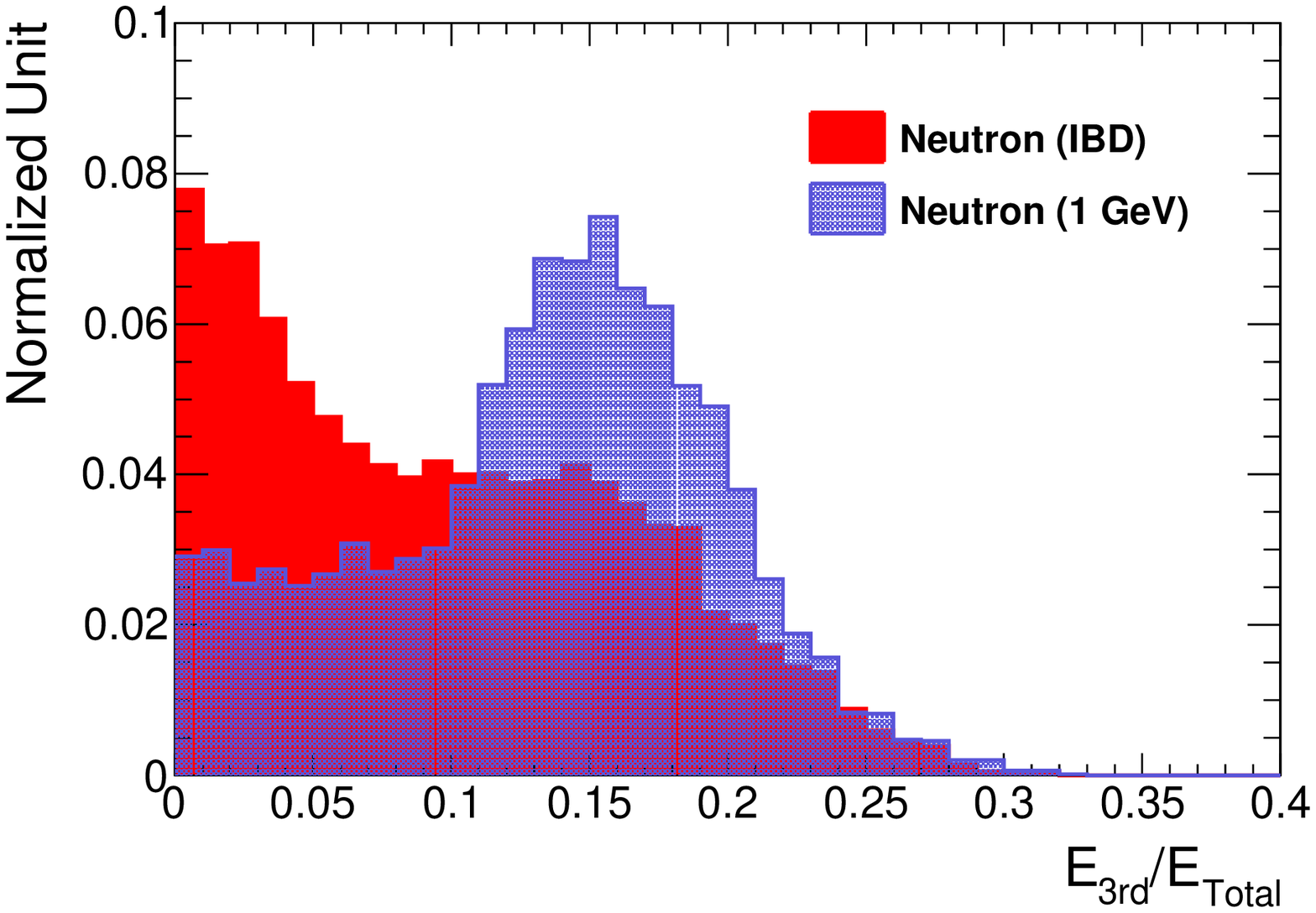}
  \caption{$E_{1st}/E_{Total}$, $E_{2nd}/E_{Total}$ and $E_{3rd}/E_{Total}$ distributions of neutron signal initiated by IBD and high energy cosmic neutron events passing through the whole detector.}
  \label{fig:Eratio_neutron}
\end{figure}

\begin{figure} [!h]
  \includegraphics[width=7.5 cm]{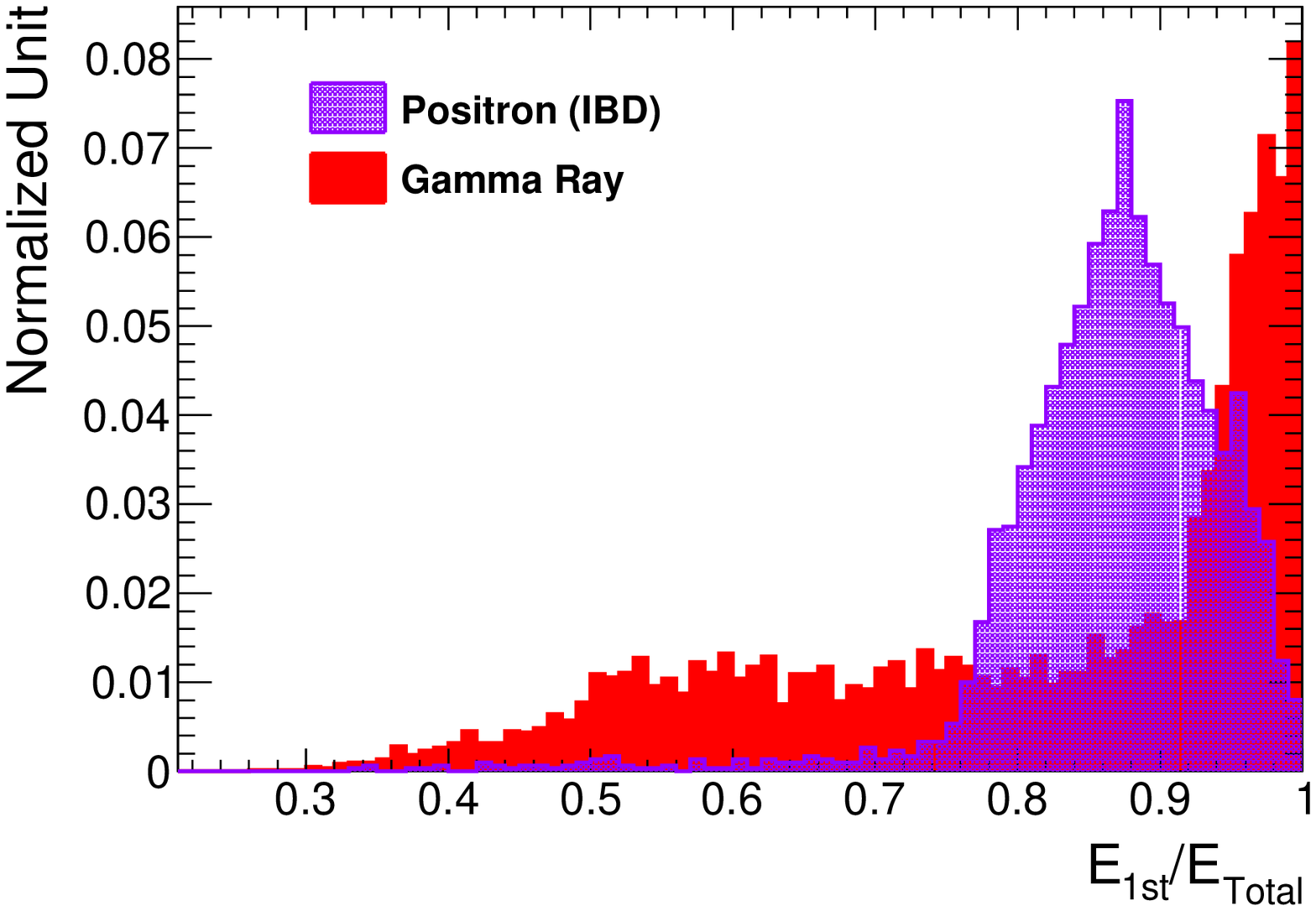}
  \includegraphics[width=7.5 cm]{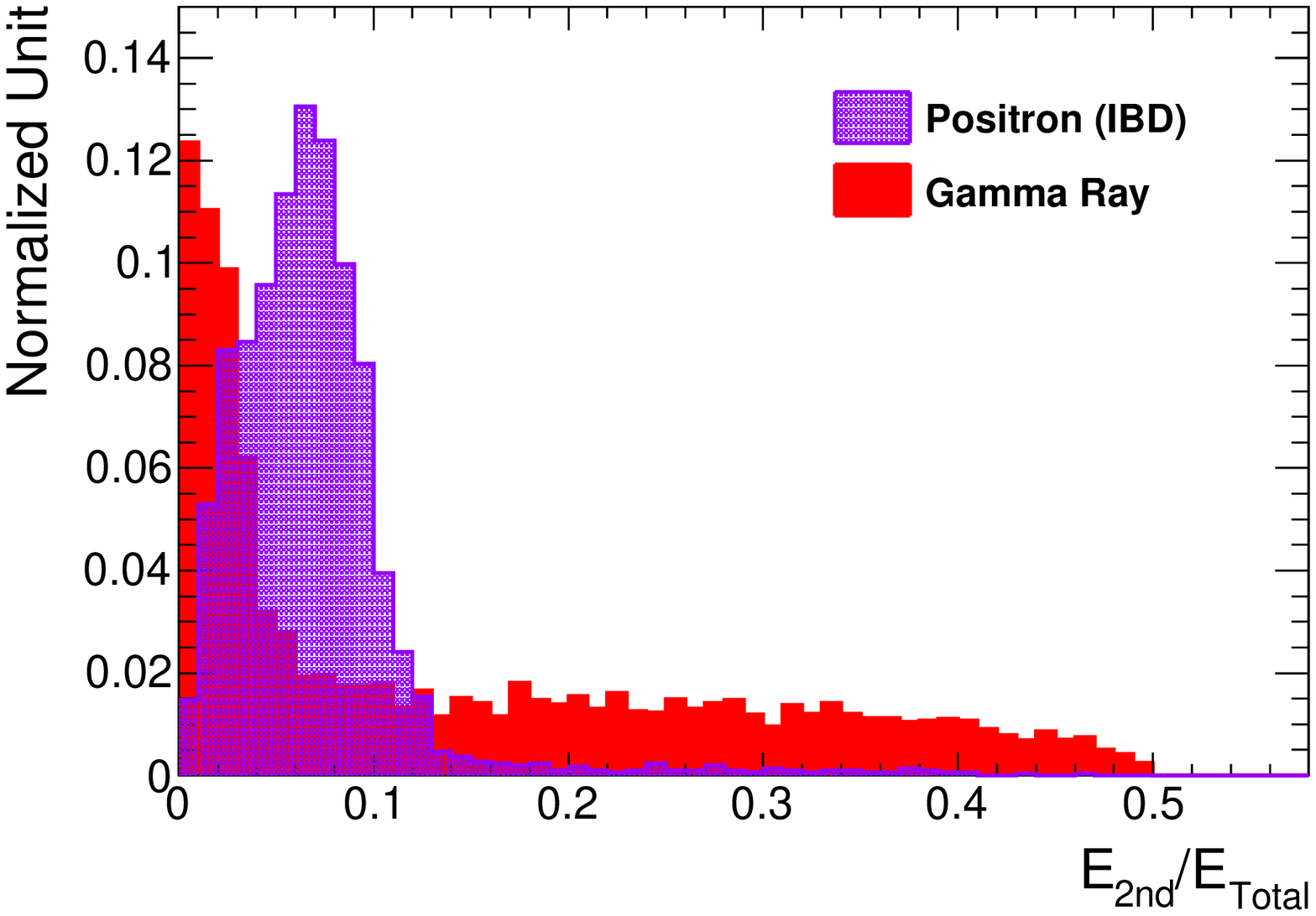}
  \includegraphics[width=7.5 cm]{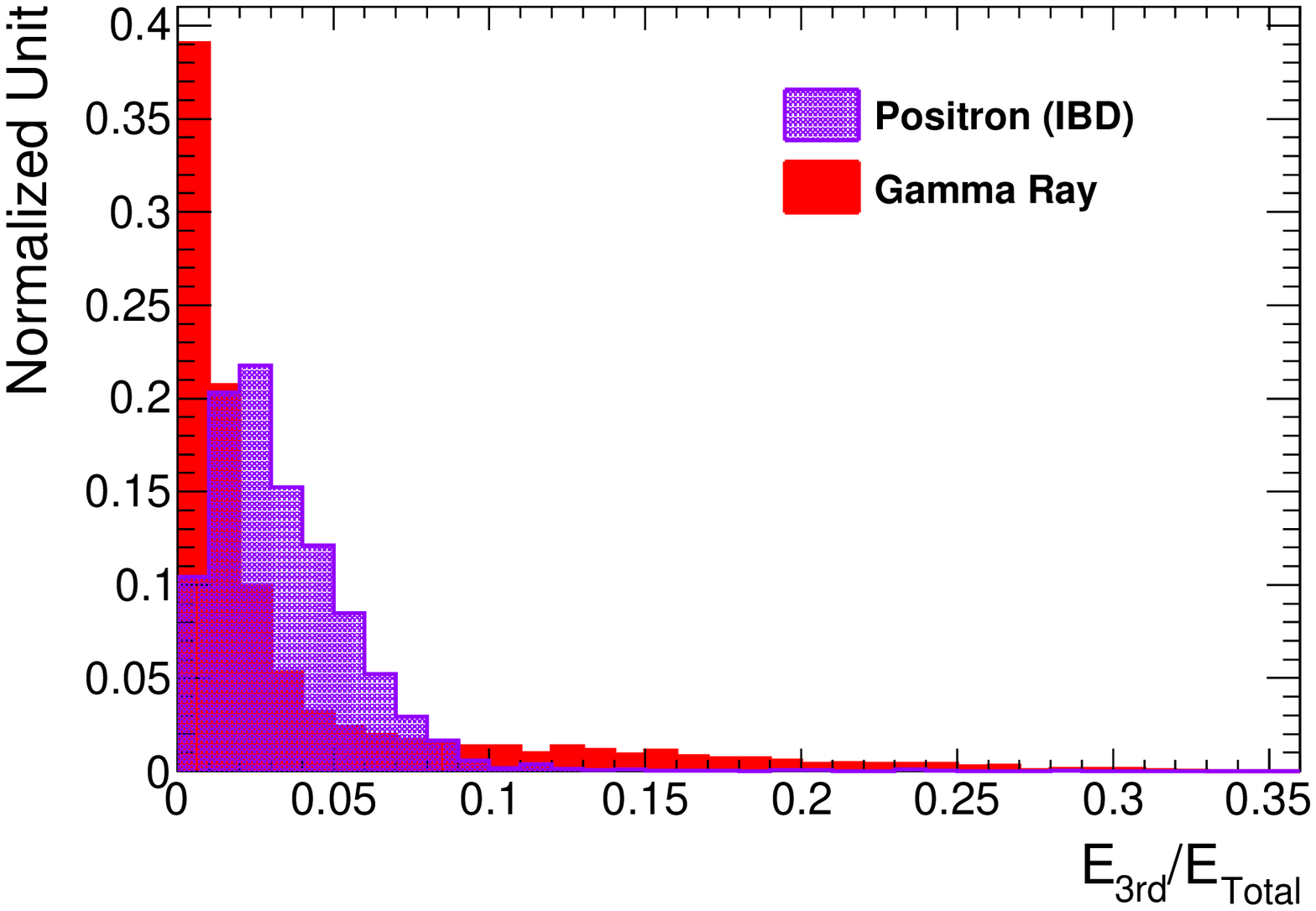}
  \caption{$E_{1st}/E_{Total}$, $E_{2nd}/E_{Total}$ and $E_{3rd}/E_{Total}$ distributions of positron signal initiated by IBD and 10 \textit{MeV} gamma ray events passing through the whole detector.}
  \label{fig:Eratio_gamma}
\end{figure}

In addition to charged cosmic rays background, high energy neutrons from cosmic shower and gamma rays above 1.022 \textit{MeV}, which can produce electron-positron pair, also are considered as background. Especially, high energy neutrons can act as IBD events \cite{Kiff}. Figure~\ref{fig:Eratio_neutron} presents comparisons of the first three highest energy fractions amongst all the modules between 10 \textit{keV} neutrons that are initiated by IBD and 1 \textit{GeV} neutrons passing through the whole detector. Similarly, the comparisons of the first three highest energy fractions between 3 \textit{MeV} positrons that are initiated by IBD and 10 \textit{MeV} gamma rays passing through the whole detector are shown in figure \ref{fig:Eratio_gamma}. 

\begin{figure} [!h]
 \centering
  \includegraphics[width=6.5 cm]{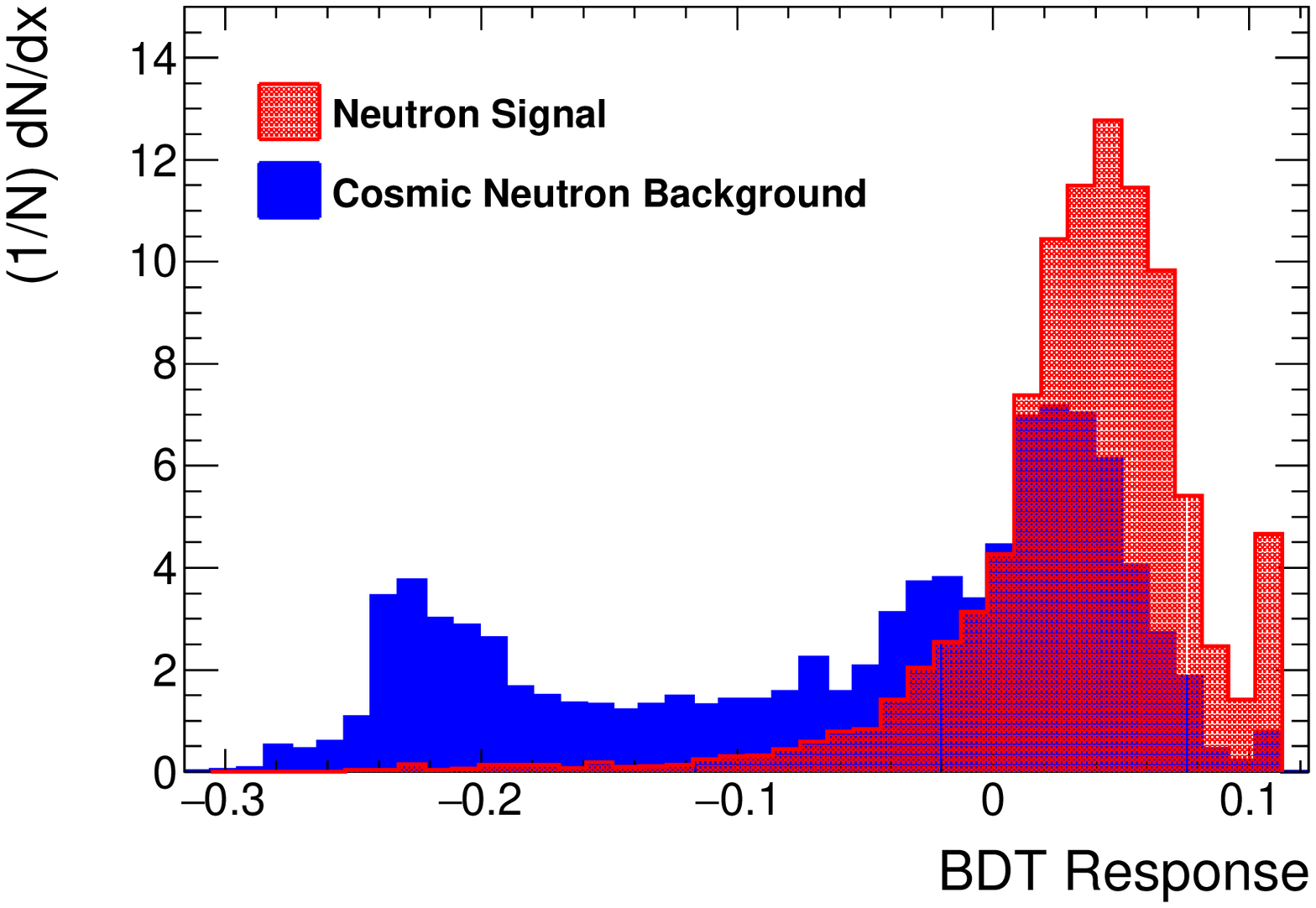}
   \includegraphics[width=6.5 cm]{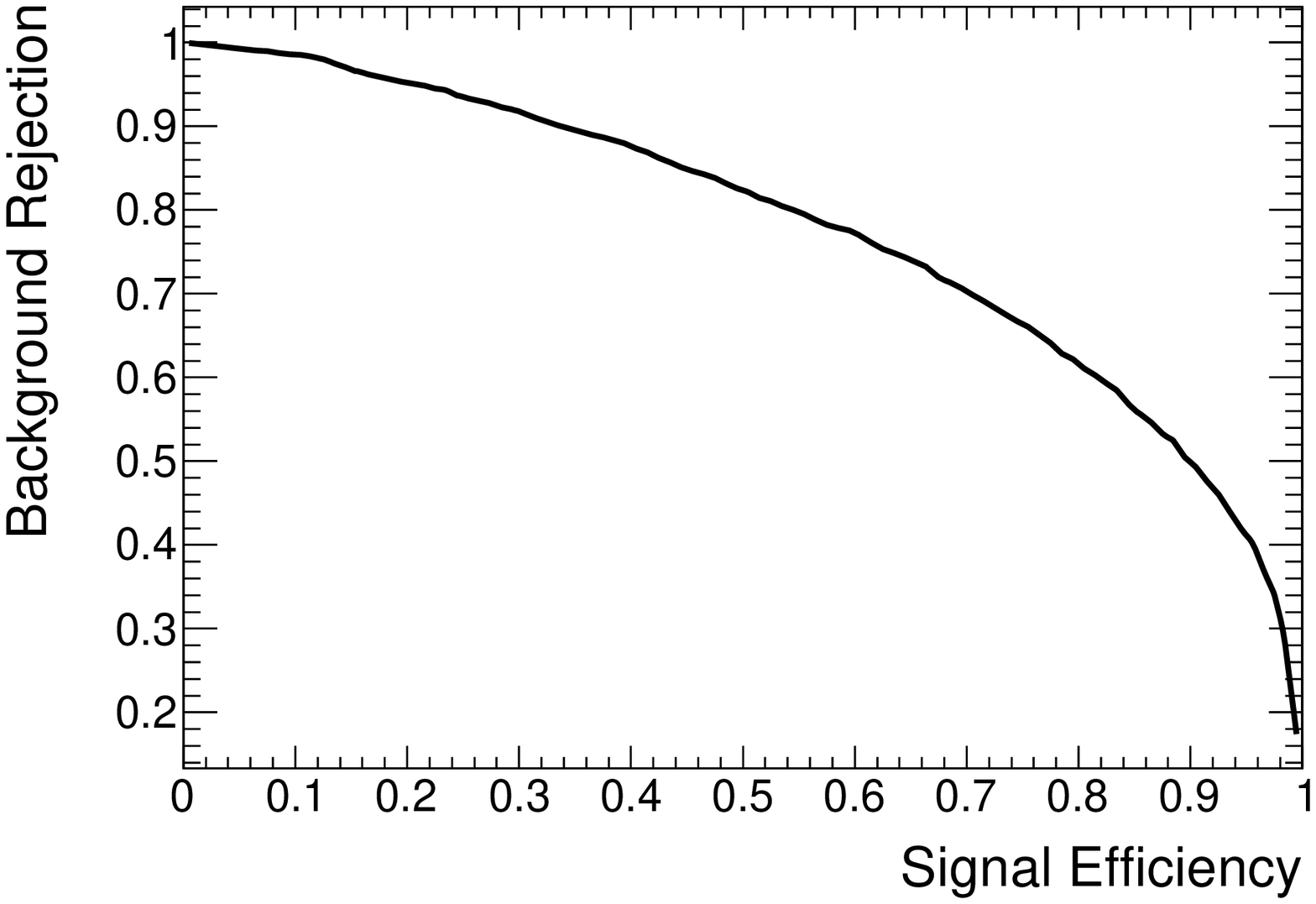}
  \caption{Boosted decision tree output distribution (left) and related ROC curve (right). Signal events are produced by neutrons resulting from IBD and the high energy cosmic neutrons passing through the whole detector are considered as background events.}
  \label{fig:BDT_neutron}
\end{figure}

\begin{figure} [!h]
 \centering
  \includegraphics[width=6.5 cm]{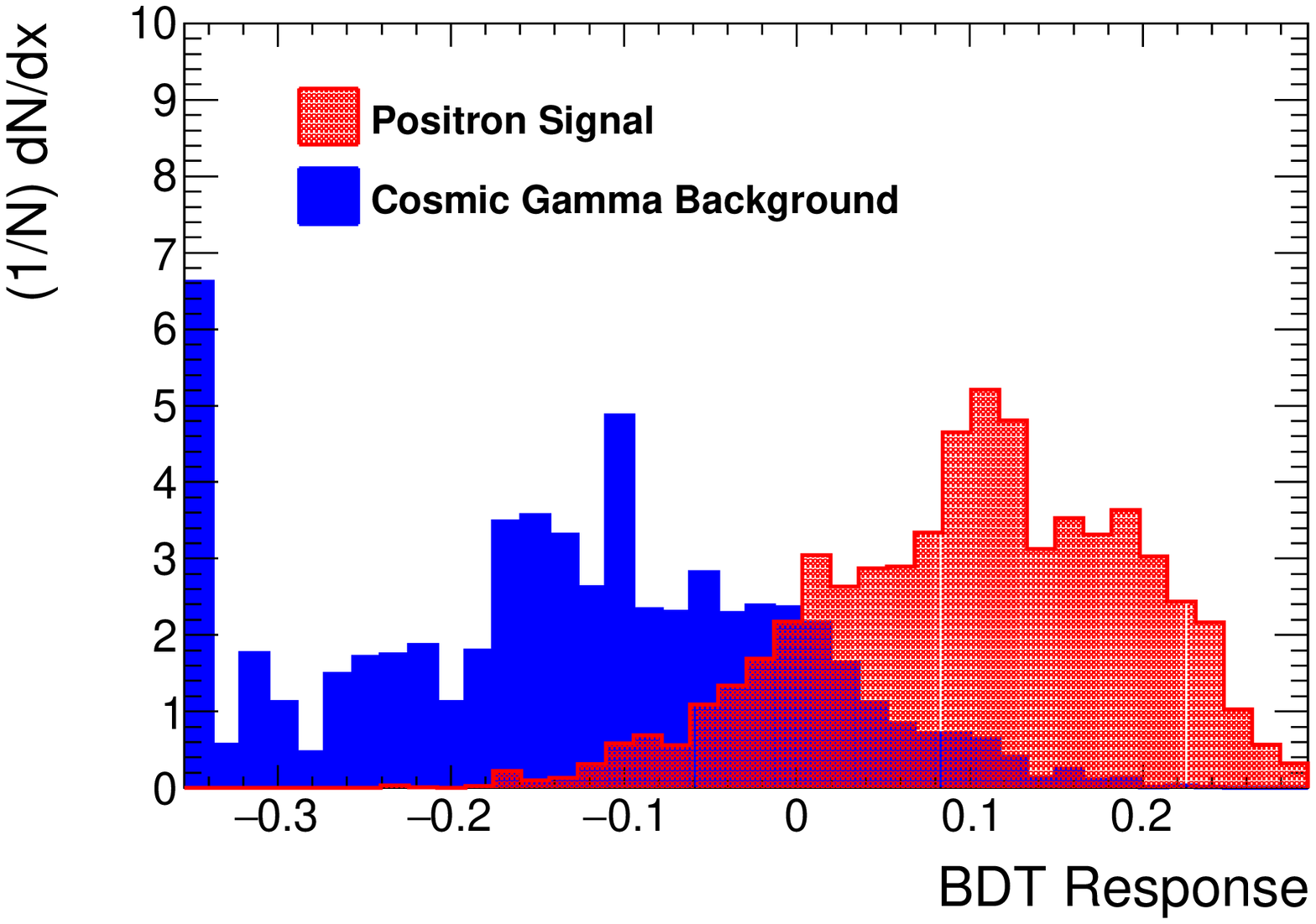}
   \includegraphics[width=6.5 cm]{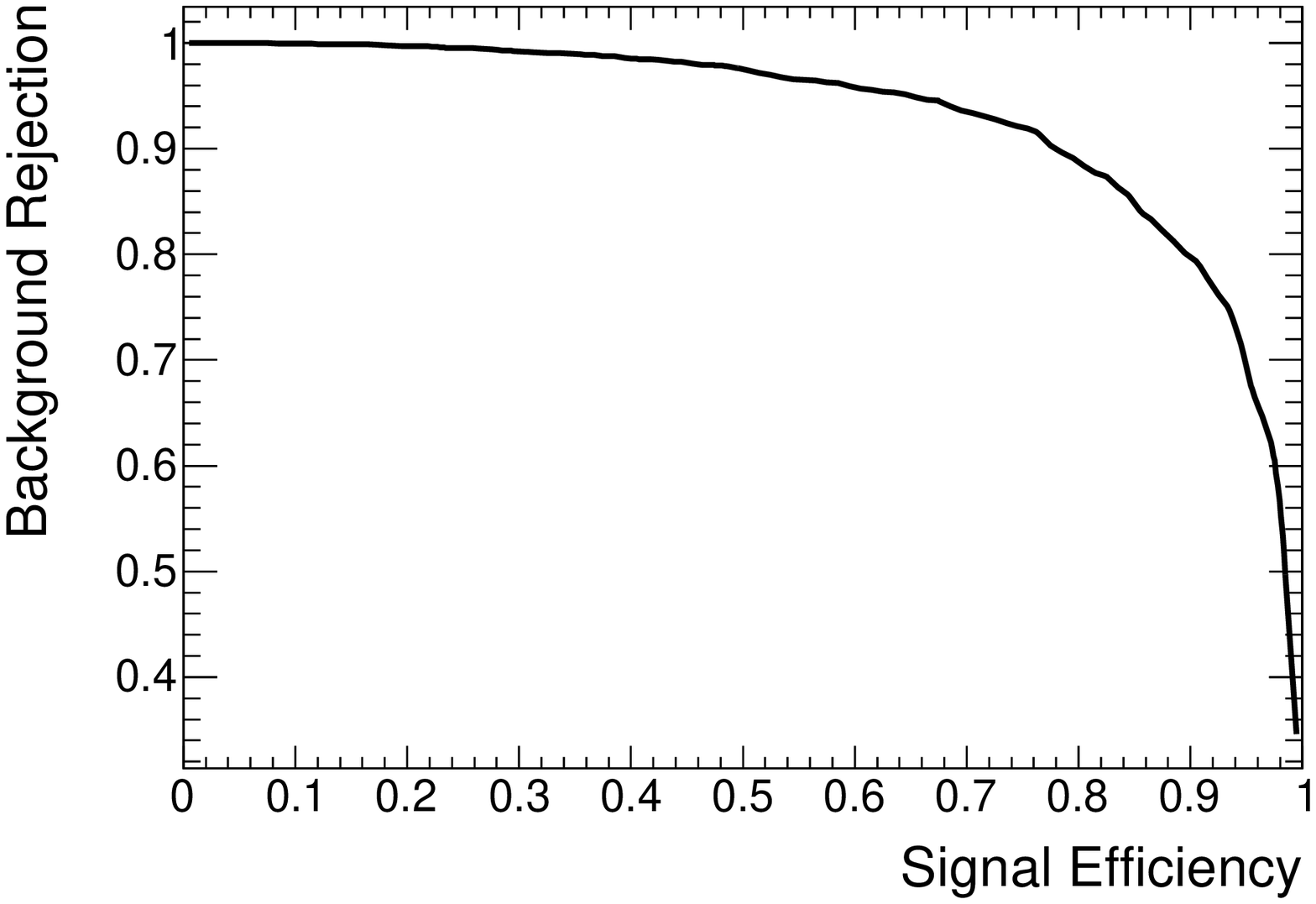}
  \caption{Boosted decision tree output distribution (left) and related ROC curve (right). Signal events are produced by positrons resulting from IBD and gamma rays with the energy of 10 \textit{MeV} passing through the whole detector are considered as background events.}
  \label{fig:BDT_gamma}
\end{figure}

In order to see how multivariate technique works for background rejection of high energy neutron from cosmic shower and gamma rays with the energy greater than 1.022 \textit{MeV}, BDT responses and ROC curves have been obtained separately. Figure \ref{fig:BDT_neutron} presents BDT response and ROC curve of IBD neutron signal while the background is only considered as high energy cosmic neutrons. Similarly,  BDT response and ROC curve of IBD positron signal, while the background is considered as gamma rays, are shown in figure \ref{fig:BDT_gamma}. These results indicate that multivariate technique might be useful for rejection of gamma rays background events with about 80\% while keeping around 90\% of antineutrino events.  In terms of high energy neutron events rejection, multivariate technique is not efficient as charged cosmic background and gamma rays background rejections. Only half of high energy cosmic neutron events could be rejected while keeping 90\% of antineutrino events.

\begin{figure} [!h]
 \centering
  \includegraphics[width=8.5 cm]{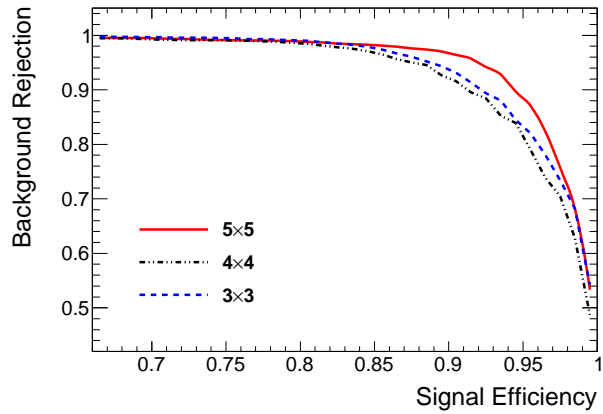}
  \caption{Comparison of ROC curves for different module configurations and numbers.}
  \label{fig:roc_comp}
\end{figure}

The effect of detector size and module numbers have also been investigated. A comparison of the ROC curves for $5\times 5$ (25 modules), $4\times 4$ (16 modules) and $3\times 3$ (9 modules) detector configurations are presented in figure~\ref{fig:roc_comp}. Since rejection of charged cosmic rays background events is the most efficient, charged cosmic particles are considered as background, while IBD neutron gives the signal. As it is seen in figure~\ref{fig:roc_comp}, no significant loss of signal-background separation is observed. This result also indicates that multivariate data analysis technique could be used for background rejection with the smaller size of the detector and with lower numbers of modules.

\clearpage

\section{Prospect}
\label{prosp}
Nuclear technology is a new area for Turkey. The first nuclear reactor in Turkey is going to start operation in 2023, in Akkuyu. Construction of additional nuclear power plants in Sinop and İgneada are being planned in the near future. These reactors will provide a  great opportunity for development of new national (and also international) neutrino physics projects. 

A kind of segmented detector made of gadolinium loaded plastic scintillators for monitoring nuclear reactors discussed in this paper must be one of the highest priority project for nuclear safety in Turkey. Such a neutrino detector could be used for monitoring of nuclear reactors and nuclear wastes  \cite{Waste} in Turkey. In addition, reactor antineutrino energy spectrum measurements could be performed with scientific purposes, and it would be the first step towards development of a new reactor neutrino oscillation experiment in Turkey. 

For that reason, it is planned to submit a project for funding to produce a demonstration module. Production and tests of gadolinium loaded plastic scintillator blocks can be done in Turkish Atomic Energy Authority, Sarayköy Nuclear Research and Training Center (SANAEM). High quality plastic scintillators have been produced in SANAEM in the past and there is necessary technological know how to synthesize new types of element loaded scintillator blocks with desired volumes and shapes.

The final design with some form of passive shielding, the construction and test are expected to take up to 2 years, before the first nuclear reactor core becomes active. 

\section{Conclusion}
\label{summary}
In this paper, design and simulation results of a segmented gadolinium-loaded polyvinyltoluene based plastic scintillator detector for nuclear reactor monitoring using antineutrino flux measurement have been presented. The detector consists of 25 identical 10$\times$10$\times$100 \textit{cm} sized of modules and it provides about 1185 antineutrino events in a day when it is located 50 \textit{m} away from 3.2 \textit{GWt} reactor core. The optimal gadolinium concentration in plastic scintillator has been found to be around 0.2\%-0.3\%, which gives prompt-delayed time difference between 4 and 50  $\mu s$ for triggering antineutrino events. 

The same simulation results have shown that background from cosmic particles can be rejected using the segmented structure of the detector without requiring any external active shielding components. In order to reject cosmic background, a multivariate data analysis technique is used taking the first three highest energy fraction in the modules as inputs. It was found that about 90\% of charged cosmic rays background rejection appears to be achievable while keeping 95\% of the antineutrino events. The results also indicate that rejection gamma rays background could be done using the multivariate technique, but this approach for high energy neutron background suppression is not efficient. This kind of a multivariate data analysis technique is considered the first time for cosmic background veto. In addition, requiring the time difference between the prompt and delayed signals, energy deposits in both signals and pulse shape analysis of the signals would increase antineutrino event detection efficiency significantly.


\section*{Acknowledgments}
I would like to thank Dr. Gokhan Unel, Dr. Erkcan Ozcan and Dr. Bora Akgun for their careful reading and useful discussion.

\end{document}